\documentclass[twocolumn,showpacs,preprintnumbers,amssymb,amsmath,pra,superscriptaddress]{revtex4}
\usepackage{amsfonts}
\usepackage{times}
\usepackage{mathrsfs}
\usepackage{graphicx}
\usepackage{dcolumn}
\usepackage{bm}
\usepackage{color}

\usepackage[colorlinks,bookmarks=false,citecolor=blue,linkcolor=red,urlcolor=blue]{hyperref}

\begin{document}
\title{Quantum Information Approach to Bose-Einstein\\ Condensate in a Tilted Double-Well System}
\author{Zhao Liu}
\author{Hongli Guo}
\email{hlguophys@iphy.ac.cn}
\author{Shu Chen}
\author{Heng Fan}
\affiliation{Institute of Physics, Chinese Academy of Sciences, Beijing, 100190, China}

\date{\today}

\begin{abstract}
We study the ground state properties of bosons in a tilted double-well system.
We use fidelity susceptibility to identify the possible ground
state transitions under different tilt values. For a very small tilt (for example $10^{-10}$),
two transitions are found. For a moderate tilt (for example $10^{-3}$), only one
transition is found. For a large tilt (for example $10^{-1}$), no transition is found.
We explain this by analyzing the spectrum of the ground state.
The quantum discord and total correlation of the ground state under different tilts
are also calculated to indicate those transitions. In the transition region,
both quantities have peaks decaying exponentially with particle number $N$.
This means for a finite-size system the transition region cannot be explained by the
mean-field theory, but in the large-$N$ limit it can be.
\end{abstract}

\maketitle


{\label{sec:level1}}
\section{Introduction}
The many-body quantum states in ultracold gases have been
studied with a high interest, because there are many parameters that
can be adjusted in experiments to control the static state as well
as the dynamics of the system. As a paradigm model, Bose-Einstein
condensate (BEC) in a double-well system provides a useful setup to tackle
the properties of quantum systems. By loading ultracold
atoms in double wells, one can study fundamental quantum mechanical
effects and many important quantum many-body phenomena, for example
interferometry \cite{Andrews}, quantum information processing
\cite{PRA73033605}, quantum phase transition \cite{nature41539},
quantum superposition state \cite{PRL87180402}, Josephson
oscillations and  nonlinear self-trapping form of dynamics
\cite{oscillations}.

Recently, it was found that in some ultracold gas systems (such as rotating
BEC and BEC in double wells with very small tilt), the ground state
transition cannot be described by the mean-field theory, although the ground states
before and after transition are consistent with the mean-field description very well
\cite{DNMJ,DNM,Zliu,AAK,BJD}. This is because in the transition region,
the ground state is no longer a product of single-particle states but a
strongly-correlated entangled state. How to characterize such states
is under intensive study. Moreover, in Refs. \cite{DNMJ,DNM,Zliu,AAK,BJD}, the system
sizes under study are not too large. The property of quantum correlation in the transition region
for larger system is an interesting problem.

The appearance of quantum correlation in the ground state in the
transition region makes it reasonable to use some tools borrowed from quantum information theory
to investigate the transition. In this article, our goal is to use fidelity susceptibility and
quantum correlation to study the ground state transition of BEC in double wells with
an arbitrary tilt. In Sec. \ref{model}, we introduce the model and give
the prediction of the ground state transition according to the semiclassical picture.
In Sec. \ref{fs}, we calculate the fidelity susceptibility which can precisely
locate the critical point of a possibly unknown
quantum transition \cite{fidelity}. Different behaviors of the fidelity susceptibility
under different tilts are found. In Sec. \ref{qd}, enlightened by the fact that
entanglement can also show a rather interesting behavior at the critical point of a quantum
transition \cite{entanglementcorrelation}, we calculate two quantities of quantum correlation:
total correlation and quantum discord, both of which can exhibit signatures of the quantum
transitions \cite{PRA80022108}. Moreover, quantum discord can appear even when
entanglement is absent \cite{PRL88017901,PRL101200501,PRA77042303,Vedral,PRA80024103} so it is a more suitable
quantity than entanglement to characterize the quantumness of the correlation.
We find that both total correlation and quantum discord are nonzero in the transition
region, but their values decrease with the particle number. By doing a
finite-size analysis, an exponential decay of them with the particle number is found.
This means although for a small system the transition is dominated by the quantum correlation,
for a very large system no quantum correlation exists during the transition.
A brief summary is given in Sec. \ref{summary}.

\section{model}
\label{model}
The single level Bose-Hubbard Hamiltonian for $N$ atoms in a double-well
system can be written as
\begin{eqnarray}
\mathcal{H}
=&-&J(\hat{a}_{L}^{\dag}\hat{a}_{R}+\hat{a}_{R}^{\dag}\hat{a}_{L})-
U[\hat{n}_{L}(\hat{n}_{L}-1)+\hat{n}_{R}(\hat{n}_{R}-1)]\nonumber\\
&-&V_{0}(\hat{n}_{L}-\hat{n}_{R}), \label{H}
\end{eqnarray}
where $\hat{a}_{i}^{\dag}$ ($\hat{a}_{i}$) creates (annihilates) a boson
in the $i$-th well ($i=L,R$), $\hat{n}_{i}=\hat{a}_{i}^{\dag}\hat{a}_{i}$,
$J$ is the tunneling energy and $U$ is the on-site interaction [a
positive (negative) $U$ corresponds to attractive (repulsive) atom-atom interaction]. $V_{0}$
is the tilt which can destroy the left-right symmetry and is
non-zero in real experiments. We set $J=1$ for convenience and only consider
$U>0$ in this paper. The above Hamiltonian can be diagonalized in
the $(N+1)$-dimensional Fock space spanned by $|n_{L},n_{R}=N-n_L\rangle$.
The dynamics of the system is controlled by the parameter $\lambda\equiv NU/J$. As $\lambda$
passes from the weak region to the fermionization limit, the
dynamics of these atoms which are initially prepared mostly in one
well, will change from Josephson oscillation (simply tunneling back
and forth between two potential wells) to self-trapping above a
critical interaction strength \cite{oscillations}.
Moreover, the static properties of the system, such as the ground state, are also closely related to $\lambda$.
We can replace the operators $\hat{a}_{i}$ with $c$ numbers $a_i=\sqrt{n_i}e^{\textrm{i}\phi_i}$
in Eq. (\ref{H}) to obtain a semiclassical Hamiltonian
\begin{eqnarray}
\frac{H}{N}=-\sqrt{1-z^2}\cos{\phi}-\frac{\lambda}{4N}(Nz^2+N-2)-V_0z,
\label{cH}
\end{eqnarray}
where $\phi=\phi_L-\phi_R$ and $z=(n_L-n_R)/N$ characterizing the imbalance.
To minimize the energy, it is obvious that $\phi$ should be zero. Then, for
each $V_0$ and $\lambda$, we can find the position $z_{\textrm{min}}$ of the
local minimum of Eq. (\ref{cH}) by solving $\frac{\partial}{\partial z}(\frac{H}{N})|_{\phi=0}=0$.
For $V_0\lesssim10^{-3}$, at a critical $\lambda\thickapprox2$, $z_{\textrm{min}}$ increases from 0
abruptly, giving
a hint of quantum transition of the ground state.

\section{fidelity susceptibility}
\label{fs}
In this section, by diagonalizing the Hamiltonian (\ref{H}), we use fidelity susceptibility to show the quantum
transition of the ground state predicted by the semiclassical Hamiltonian (\ref{cH}).
The ground state fidelity susceptibility is defined as
\begin{eqnarray}
\chi(\lambda)=-\lim_{\delta \lambda\rightarrow0}\frac{2 \ln
F}{\delta
\lambda^{2}}=\sum_{n\neq0}\frac{|\langle\Psi_{n}(\lambda)|\frac{\partial\mathcal
{H}}{\partial\lambda}|\Psi_{0}(\lambda)\rangle|^{2}}{[E_{n}(\lambda)-E_{0}(\lambda)]},\nonumber
\end{eqnarray}
where $\Psi_{0}(\lambda)$ [$\Psi_{n}(\lambda)$] is the ground
(excited) state of $\mathcal {H}$, $E_{n}(\lambda)$ and
[$E_{0}(\lambda)$] is the corresponding ground (excited) energy.
Here we suppose that $\mathcal {H}$ has a non-degenerate ground
state. In our system, $\frac{\partial\mathcal
{H}}{\partial\lambda}=-\frac{1}{N}[\hat{n}_{L}(\hat{n}_{L}-1)+\hat{n}_{R}(\hat{n}_{R}-1)]$.
When $V_{0}=0$, the ground states are degenerate for some values of $\lambda$. To break this
degeneracy, we add an non-zero tilt and then we can use fidelity
susceptibility to study the quantum transition of the ground
state.

\begin{figure}
\centerline{\includegraphics[width=\linewidth]{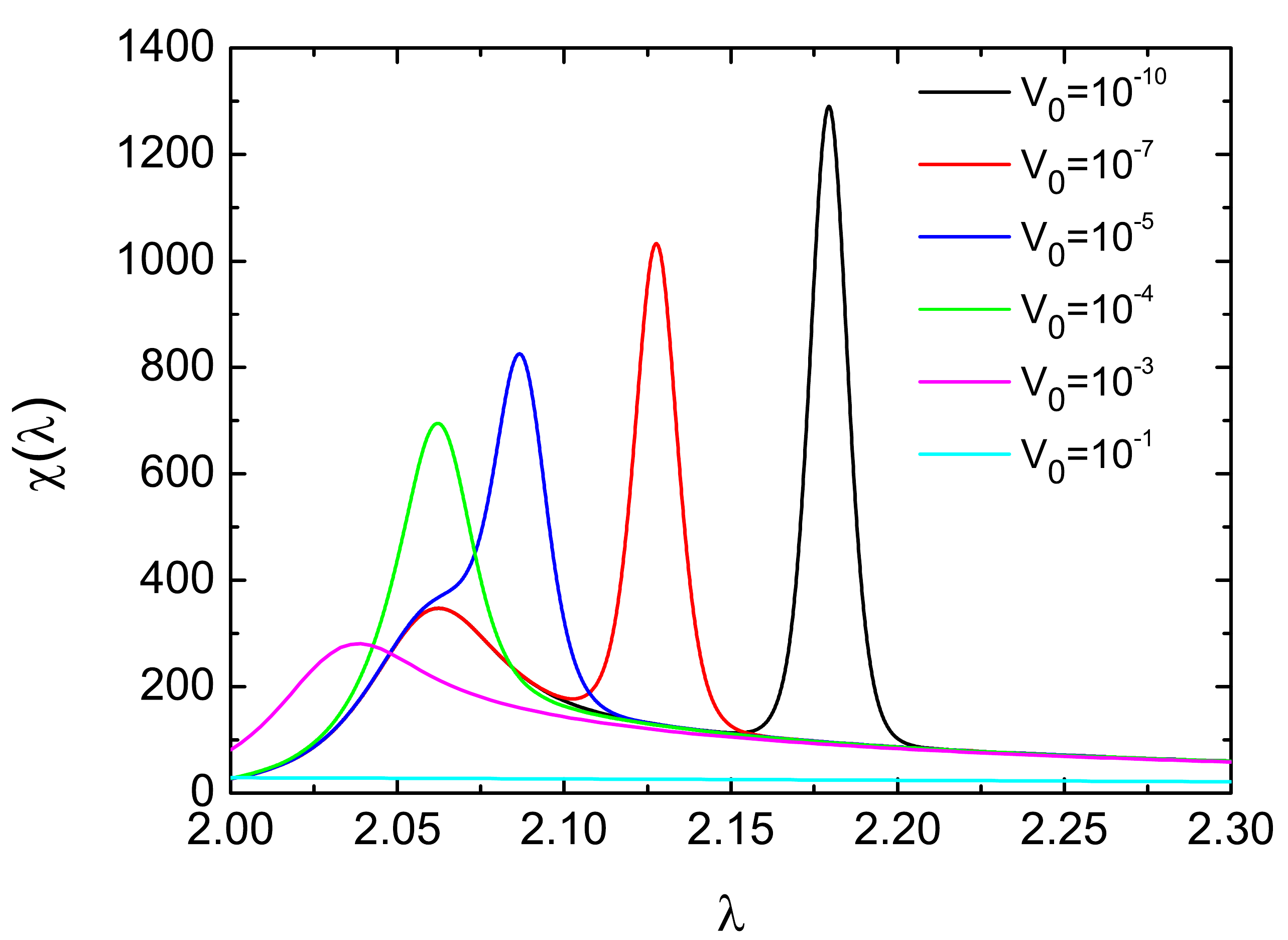}}
\caption{(color online) The fidelity susceptibility $\chi(\lambda)$
for $N=800$ under tilt $V_{0}=10^{-10}$, $10^{-7}$, $10^{-4}$,
$10^{-3}$ and $10^{-1}$. One can see a clear transition from double peaks to single peak
around $V_0=10^{-5}$.}\label{chiV0}
\end{figure}

In Fig. \ref{chiV0}, we fix the particle number $N$ to show the relation between
$\chi(\lambda)$ and $V_0$. We find that under a small $V_0$,
$\chi(\lambda)$ has two peaks. With the increase of $V_0$, the position of the
left peak does not change but the right peak moves left towards smaller $\lambda$.
When $V_0$ is moderate (for example when $V_0=10^{-4}$), only one peak remains and
continues moving left. When $V_0$ is large enough (for example when $V_0=10^{-1}$),
no peak remains.

\begin{figure*}
\centerline{\includegraphics[width=\linewidth]{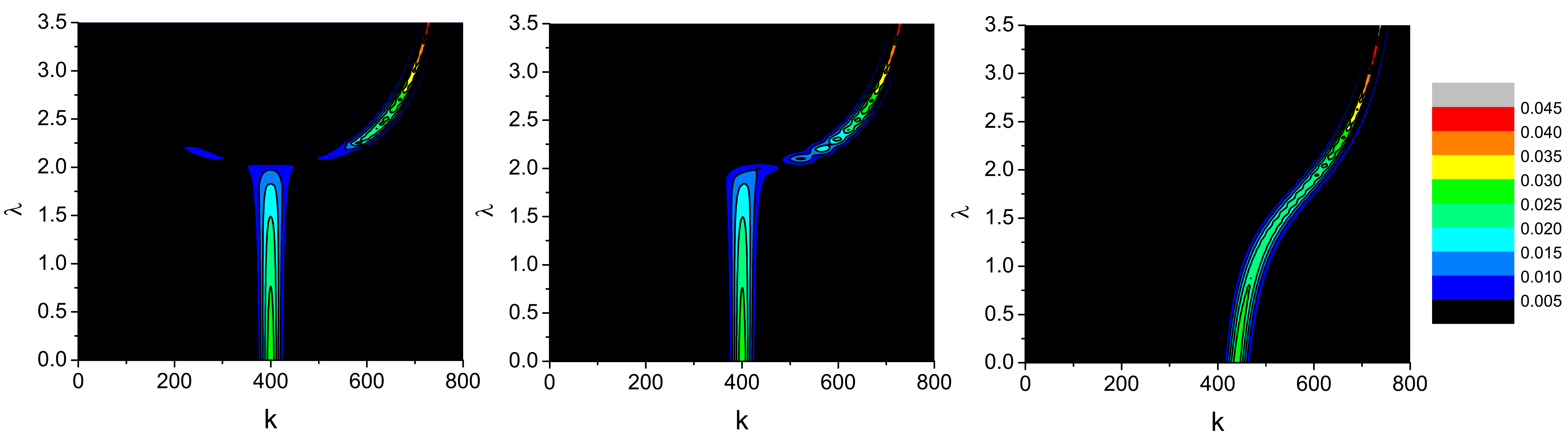}}
\caption{(color online) The spectrum $|c_k|^2$ as a function of $k$
for different $\lambda$ under different tilts $V_{0}=10^{-10}$
(left), $V_0=10^{-3}$ (middle) and $V_0=10^{-1}$ (right). We choose $N=800$
in this figure.}\label{ck}
\end{figure*}

The behavior of $\chi(\lambda)$ can be understood from the analysis
of the ground state. The ground state can be expanded as
$|\Psi_0\rangle=\sum_{k=0}^N c_k |k,N-k\rangle$, where $n_L=k$ and
$n_R=N-k$. Through studying the spectrum $|c_k|^2$ as a function of
$k$, we can know the configuration of particles in the two wells.
Fig. \ref{ck} shows the spectra under different tilts
$V_{0}=10^{-10}$, $10^{-3}$, and $10^{-1}$. Under $V_{0}=10^{-10}$
[Fig. \ref{ck}(a)], when $\lambda\lesssim2.06$, $|c_k|^2$ is
symmetric and has a peak at $k=N/2$, which is consistent with
the knowledge that the ground state is a binomial state
$|\Psi_0\rangle=\frac{1}{\sqrt{2^n}}\sum_{k=0}^N
\sqrt{\frac{n!}{k!(n-k)!}}|k,N-k\rangle$ at $\lambda=0$ under
$V_0=0$. When $2.06\lesssim \lambda\lesssim2.22$, $|c_k|^2$ is still
symmetric but has two peaks. This means the ground state becomes a
cat-like state. When $\lambda \gtrsim 2.22$, $|c_k|^2$ only has
one peak which moves towards $k=N$, meaning all particles tend to
locate in one well and self-trapping occurs. Therefore, at small
enough tilt, there are three phases, reflected by two peaks of
fidelity susceptibility. If we increase the tilt, the cat-like
region becomes smaller and finally vanishes after the tilt is increased
to an appropriate value, for example $V_{0}=10^{-3}$ in Fig.
\ref{ck}(b). At this value of the tilt, the ground state will change
from the binomial configuration to the self-trapping directly at
some critical $\lambda$, reflected by the single peak of
$\chi(\lambda)$. For a very large tilt, $|c_k|^2$ is
not symmetric in the whole region [Fig. \ref{ck}(c)] and no ground state transition
appears.

\begin{figure*}
\centerline{\includegraphics[width=\linewidth]{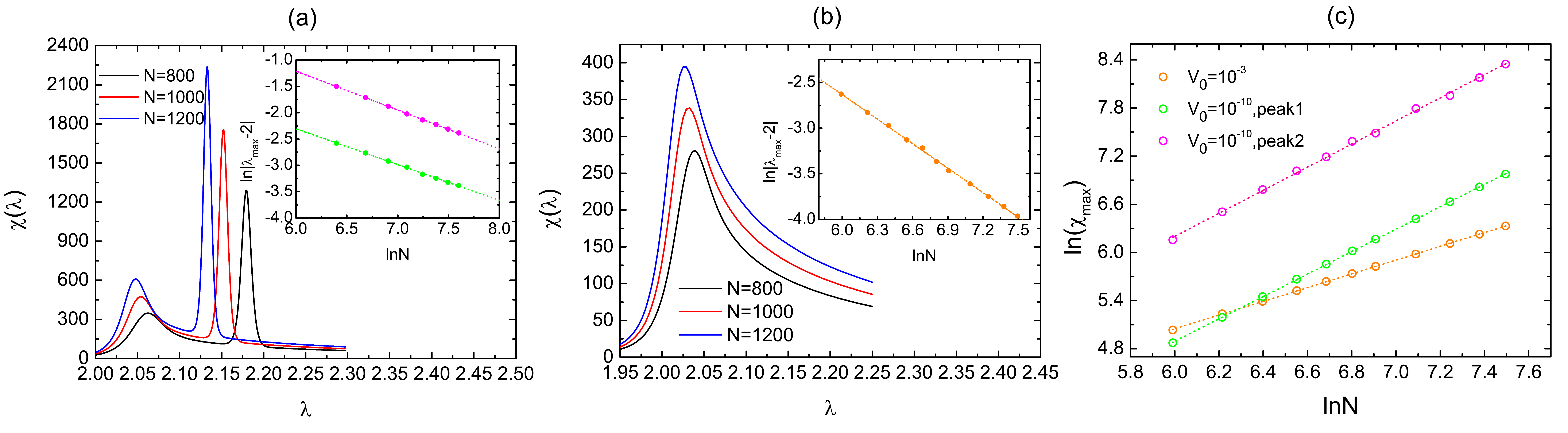}}
\caption{(color online) (a) The fidelity susceptibility $\chi(\lambda)$
for different particle numbers $N=$800, 1000 and 1200 under a small
tilt $V_{0}=10^{-10}$. The inset shows the finite-size scaling
analysis of the position $\lambda_{\max}$ of each peak.
One can see that for both peaks, $\ln|\lambda_{\max}-2|\propto\ln N$ with a negative slope
[green dots for the left peak (peak 1) and magenta dots for the right peak (peak2)].
Therefore in the large-$N$ limit, both peaks are at the same position
$\lambda_{\max}=2$. (b) The fidelity susceptibility $\chi(\lambda)$
for different particle numbers $N=$800, 1000 and 1200 under a moderate
tilt $V_{0}=10^{-3}$. The inset shows the finite-size scaling
analysis of the position $\lambda_{\max}$ of the single peak. Here one can also find that
$\ln|\lambda_{\max}-2|\propto\ln N$ with a negative slope and in the large-$N$ limit $\lambda_{\max}=2$.
(c) The finite-size scaling analysis of $\chi(\lambda_{\max})$ under small and moderate tilt, which diverges
exponentially with $N$.}
\label{chiN}
\end{figure*}

So far our discussion is based on a fixed particle number $N$. Now we need
to fix the tilt and enlarge $N$ to see what happens in the large-$N$ limit.
The height of peaks in both double-peak and single-peak region diverge with $N$ exponentially
as predicted by the fidelity susceptibility theory [Fig. \ref{chiN}(c)].
In the double-peak region [Fig. \ref{chiN}(a)], when we increase $N$, the distance
between the two peaks becomes smaller. A finite-size analysis of the positions $\lambda_{\textrm{max}}$ of both
peaks shows that $|\lambda_{\textrm{max}}-2|\propto N^{-d_{p}}$ with $d_p\approx0.6799$
for the left peak and $d_p\approx0.738$ for the right peak. Therefore, $\lambda_{\textrm{max}}$
of both peaks will tend to 2 when $N\rightarrow\infty$, being consistent with the
prediction of Eq. (\ref{cH}). Similarly, in the single-peak region [Fig. \ref{chiN}(b)],
$|\lambda_{\textrm{max}}-2|$ is also proportional to $N^{-d_{p}}$ with $d_p\approx0.8941$.
We can know from these results that in the thermodynamic limit there is only one direct \emph{quantum phase
transition} from the binomial state to the self-trapping state. The double-transition from the binomial state
to the cat state then to the self-trapping state under a small tilt and for moderate $N$
is actually a \emph{crossover}. Our results confirm that the fidelity susceptibility
is useful for detecting not only quantum phase transitions but also crossovers \cite{crossover}.

\section{Quantum Discord}
\label{qd}
It is usually stated that the quantum transition of the ground state
can be indicated by some quantum information quantity, such as entanglement
of the ground state.
Here we revisit this problem by studying the correlation in the transition
region of our system. Before we discuss this, we first introduce two quantities
we use to describe the correlation: the total correlation and quantum discord.

Suppose that we have a system $AB$ composed by two subsystems $A$ and $B$.
Then we can use three density matrices $\rho_{AB}$, $\rho_A$ and
$\rho_B$ to describe the states of the whole system and the two
subsystems respectively, where
$\rho_{A(B)}=\textrm{Tr}_{B(A)}\rho_{AB}$. The joint entropy of the
whole system is defined as the von Neumann entropy of $\rho_{AB}$:
$S(\rho_{AB}) = -\textrm{Tr} (\rho_{AB} \ln \rho_{AB})$. Similarly, we can also calculate the von Neumann
entropy $S (\rho_{A(B)})$ of $\rho_{A(B)}$. If $\rho_{AB}$ is a pure
state, $S (\rho_{A(B)}) = -\textrm{Tr} (\rho_{A(B)} \ln
\rho_{A(B)})$ is called as entanglement entropy and used to quantify
the quantum entanglement between $A$ and $B$. However, if $\rho_{AB}$
is a mixed state (like the two-body reduced density matrix in our system),
$S (\rho_{A(B)})$ is not a good measure of entanglement.
The total correlation
(the quantum mutual information) between $A$ and $B$ is given by
$\mathcal {I}_{AB}=
S(\rho_{A})-S(\rho_{A}|\rho_{B})$, where $S(\rho_{A}|\rho_{B})=
S(\rho_{AB})- S(\rho_{B})$.
Generally speaking, a bipartite quantum state $\rho_{AB}$ has both classical and quantum characteristics. So
we can divide the the total correlation $\mathcal {I}_{AB}$ into two parts: the quantum part
and the classical part.

The classical part is defined as the maximum
information about one subsystem that can be obtained by performing
measurements on the other subsystem. Let us consider a measurement performed only on subsystem $B$.
This measurement can be described by a complete set of projectors
$\{\mathcal {M}_{k}\}$ where $\mathcal {M}_{k}\geq0$ and
$\sum_k\mathcal {M}_{k}=\mathbb{I}_{B}$. The state of system $AB$ after the
application of $\mathcal {M}_{k}$ becomes
$\rho_{AB}^k=\frac{1}{p_{k}}(\mathbb{I}_A\otimes \mathcal {M}_{k})\rho_{AB}
(\mathbb{I}_A\otimes \mathcal {M}_{k})$ with $p_{k}=\textrm{Tr}[(\mathbb{I}_A\otimes
\mathcal {M}_{k})\rho_{AB}(\mathbb{I}_A\otimes \mathcal {M}_{k})]$. According
to the definition, the classical correlation can be obtained as
$\mathcal {C}_{AB}=S(\rho_{A})-\min_{\{\mathcal
{M}_{k}\}}\sum_{k}p_kS(\rho_{A}^k)$, where
$\rho_{A}=\textrm{Tr}_B\rho_{AB}$ and
$\rho_{A}^k=\textrm{Tr}_B\rho_{AB}^k$.
The quantum component of the
correlation between two systems can be regarded as the difference
between the total correlation and the classical correlation.
This quantity is what we call quantum discord
$\mathcal {D}_{AB}=\mathcal {I}_{AB}-\mathcal {C}_{AB}$.
It is interesting that some non-entangled states can also have nonzero quantum discord \cite{PRL88017901,PRL101200501,PRA77042303,Vedral,PRA80024103}, revealing that quantum
discord is more suitable to capture the quantumness of correlation
than entanglement. For pure states, the discord
reduces exactly to the entanglement entropy.

Now we consider quantum discord and the total correlation
between two particles in our double-well system (See Ref. \cite{mode entanglement}
for the discussion of the entanglement between identical particles). Because all particles
are identical qubits, we have
$\mathcal {I}_{AB}=2S_1-S_2$,
where $S_{1} (S_{2})$ is the von Neumann entropy of one- (two-) particle reduced density matrix.
Here the one-particle and two-particle reduced density matrices
are defined as $(\rho_1)_{ij}=\frac{1}{N}\langle a_j^{\dagger}a_i\rangle$
and $(\rho_2)_{ij,kl}=\frac{1}{N(N-1)}\langle a_k^{\dagger}a_l^{\dagger}a_j a_i\rangle$ respectively,
where $i,j,k,l\in\{L,R\}$ and the average is made under the ground state.
For qubits, each complete set of projectors contains two elements
labeled by two parameters $\theta\in[0,\pi]$ and $\varphi\in[0,2\pi]$, such that
\begin{eqnarray}
\mathcal{M}_1(\theta,\varphi)=\left(
\begin{array}{cccc}
\cos^2\frac{\theta}{2}&\sin\frac{\theta}{2}\cos\frac{\theta}{2}e^{-\textrm{i}\varphi}\\
\sin\frac{\theta}{2}\cos\frac{\theta}{2}e^{\textrm{i}\varphi}&\sin^2\frac{\theta}{2}
\end{array}\right),\nonumber\\
\mathcal{M}_2(\theta,\varphi)=\left(
\begin{array}{cccc}
\sin^2\frac{\theta}{2}&-\sin\frac{\theta}{2}\cos\frac{\theta}{2}e^{-\textrm{i}\varphi}\\
-\sin\frac{\theta}{2}\cos\frac{\theta}{2}e^{\textrm{i}\varphi}&\cos^2\frac{\theta}{2}
\end{array}\right).\nonumber
\end{eqnarray}
One can note that actually we can write
$\mathcal{M}_i(\theta,\varphi)=|\Phi_i(\theta,\varphi)\rangle\langle\Phi_i(\theta,\varphi)|$ ($i=1,2$)
with $|\Phi_1(\theta,\varphi)\rangle=\left(
\begin{array}{cccc}
\cos\frac{\theta}{2}\\
\sin\frac{\theta}{2}e^{\textrm{i}\varphi}
\end{array}\right)$ and $|\Phi_2(\theta,\varphi)\rangle=\left(
\begin{array}{cccc}
\sin\frac{\theta}{2}\\
-\cos\frac{\theta}{2}e^{\textrm{i}\varphi}
\end{array}\right)$.
Then the quantum discord can be expressed as
\begin{eqnarray}
\mathcal {D}_{AB}=S_1-S_2+\min_{\{\theta,\varphi\}}\sum_{k=1}^2p_k(\theta,\varphi)
S[\rho_{2}^k(\theta,\varphi)],\nonumber
\end{eqnarray}
where $p_k(\theta,\varphi)=\textrm{Tr}[(\mathbb{I}\otimes
\mathcal {M}_{k}(\theta,\varphi))\rho_{2}(\mathbb{I}\otimes \mathcal {M}_{k}(\theta,\varphi))]$
and $\rho_{2}^k(\theta,\varphi)=\frac{1}{p_k(\theta,\varphi)}(\mathbb{I}\otimes
\mathcal {M}_{k}(\theta,\varphi))\rho_{2}(\mathbb{I}\otimes \mathcal {M}_{k}(\theta,\varphi))$.
Although in some cases one can
obtain an analytical expression for $\mathcal {D}_{AB}$ \cite{PRA77042303}, we have to
do a numerical calculation here. We divide the domains of $\theta$ ($[0,\pi]$) and $\phi$
($[0,2\pi]$) into 100 equal intervals respectively and search the minimization.

\begin{figure}
\centerline{\includegraphics[width=\linewidth]{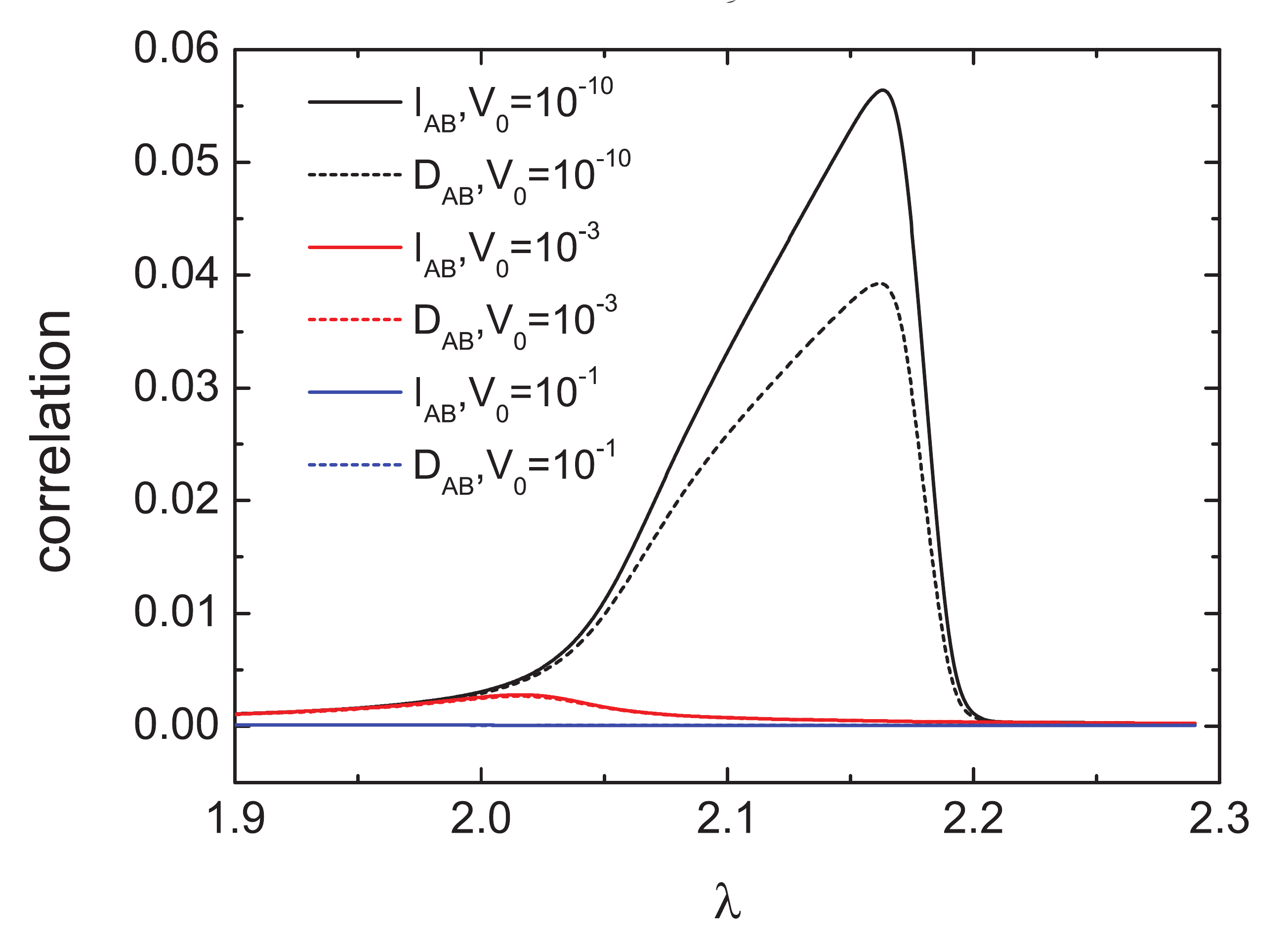}}
\caption{(color online) The quantum discord $\mathcal {D}_{AB}$
and total correlation $\mathcal {I}_{AB}$ for $N=800$ under different
tilts $V_0=10^{-10}$, $10^{-3}$ and $10^{-1}$. One can see that
with the increase of $V_0$, both $\mathcal {D}_{AB}$ and $\mathcal {I}_{AB}$
decrease.}
\label{DV0}
\end{figure}

Similar with what we did in Sec. \ref{fs}, we first fix the particle
number $N$ to study the relation between correlations and $V_0$ (Fig. \ref{DV0}).
We find that $\mathcal {D}_{AB}$ and $\mathcal {I}_{AB}$ have similar
behaviors. Under a small tilt $V_0=10^{-10}$, each correlation
in the transition region is remarkably larger than that out of the transition
region, which means this transition cannot be described by the mean field theory
for this system size. However, for each correlation we only observe one peak,
whose position is near the position of the right peak of
$\chi(\lambda)$ [Fig. \ref{chiN}(a)]. If we increase the tilt to $V_0=10^{-3}$, both
$\mathcal {D}_{AB}$ and $\mathcal {I}_{AB}$ become smaller. However,
a peak still exists, whose position is near the position of the single peak of $\chi(\lambda)$ [Fig. \ref{chiN}(b)].
If the tilt is further increased, both $\mathcal {D}_{AB}$ and $\mathcal {I}_{AB}$
are almost zero, meaning the ground state is almost a product
state for any $\lambda$.

\begin{figure*}
\centerline{\includegraphics[width=\linewidth]{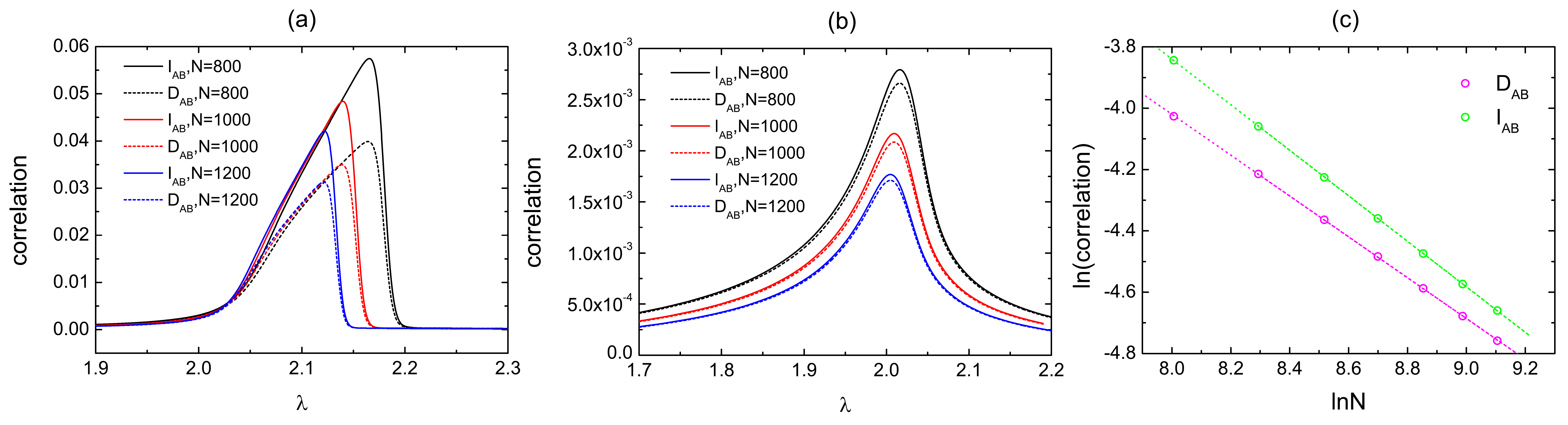}}
\caption{\label{fig:epsart} (a) The quantum discord $\mathcal {D}_{AB}$ and the total correlation
$\mathcal {I}_{AB}$ for different particle number $N=800$, 1000 and 1200
under a small tilt $V_{0}=10^{-10}$. One can find that with the increase of
$N$, both $\mathcal {D}_{AB}$ and $\mathcal {I}_{AB}$ decrease.
(b) The quantum discord $\mathcal {D}_{AB}$ and the total correlation
$\mathcal {I}_{AB}$ for different particle number $N=800$, 1000 and 1200
under a moderate tilt $V_{0}=10^{-3}$. One can also find that with the increase of
$N$, both $\mathcal {D}_{AB}$ and $\mathcal {I}_{AB}$ decrease.
(c) The finite-size scaling analysis of the peak values of both $\mathcal {D}_{AB}$ and $\mathcal {I}_{AB}$
under the small tilt $V_{0}=10^{-10}$ from $N=3000$ to $N=9000$.
One can see that $\ln\mathcal {D}_{AB}^{\max}$ ($\ln\mathcal {I}_{AB}^{\max}$) $\propto\ln N$ with a negative slope,
where $\mathcal {D}_{AB}^{\max}$ ($\mathcal {I}_{AB}^{\max}$) is the peak value
of $\mathcal {D}_{AB}$ ($\mathcal {I}_{AB}$). This means both of them decay to zero in large-$N$ limit.
A similar analysis (not shown here) demonstrates that
under a moderate $V_0=10^{-3}$, the peak values of both $\mathcal {D}_{AB}$ and $\mathcal {I}_{AB}$ also decay to zero in large-$N$ limit.}
\label{staticD}
\end{figure*}

Then, we want to know whether we can have a non-zero correlation in the large-$N$ limit.
After fixing $V_0$, we find both $\mathcal {D}_{AB}$ and $\mathcal {I}_{AB}$ decrease with $N$ [Fig. \ref{staticD}(a) and (b)].
Through a finite-size analysis for the tilt $V_0=10^{-10}$, we find the peak values of the correlations
decay exponentially as $N^{-d_c}$ ($d_c\approx0.6661$ for quantum discord and $d_c\approx0.7426$
for the total correlation). This means in the $N\rightarrow \infty$ limit, there will not be correlation
in the transition region and this transition can be described by the mean field theory.
For $V_0=10^{-3}$, a similar conclusion can also be obtained.

\section{summary}
\label{summary}
In this paper, we analyze the quantum transition of the ground state for
the single level Bose-Hubbard model in a double-well system with an arbitrary tilt. A semiclassical Hamiltonian
predicts that for a not too large tilt, this transition happens at $NU/J=2$.
We use fidelity susceptibility $\chi(\lambda)$ to identify this transition. We find that
for a small tilt, $\chi(\lambda)$ has two peaks which are at the same position in the
$N\rightarrow\infty$ limit. One peak corresponds to the transition from a binomial state
to a cat-like state and the other peak corresponds to the transition from a cat-like state
to the self-trapping state. While for a moderate tilt, only one peak of $\chi(\lambda)$ is
observed, which corresponds to the direct transition from a binomial state to self-trapping.
For a large tilt, no transition is observed in $\chi(\lambda)$.

We also use two quantities describing correlation, quantum discord and the total correlation,
to indicate the ground state transition. For a finite system size, each correlation has a
peak in the transition region (either for a small tilt or a moderate tilt), meaning the
transition cannot be described by the mean field theory. However, by doing a finite-size
analysis, we find that in the $N\rightarrow\infty$ limit, both correlations decay exponentially with
particle number to zero. It is an interesting generalization to check the behavior of
quantum correlation with system size in other systems where a ground state transition
that cannot be described by the mean-field theory exists, such as rotating BEC.

\begin{acknowledgments}
Zhao Liu thanks the financial support from the MPG---CAS Joint
Doctoral Promotion Programme (DPP) and Max-Planck Institute of
Quantum Optics. Hongli Guo thanks the financial support from the MPG---CAS Joint
Doctoral Promotion Programme (DPP) and Max-Planck Institute for the Physics of
Complex Systems. Heng Fan is supported by ``973" program
(Grant No. 2010CB922904).
\end{acknowledgments}

\emph{Note Added}-- Zhao Liu and Hongli Guo equally contributed to this work.
Hongli Guo is the corresponding author of this article.

\end{document}